\documentclass{aastex631}

\usepackage[utf8]{inputenc}
\usepackage{xcolor}
\usepackage{hyperref}
\usepackage{graphicx}

\newcommand{\myurl}[2][red]{\href{#2}{\color{#1}{#2}}}%

\begin{document}

\received{}
\revised{}
\accepted{}
\submitjournal{ApJS}

\shorttitle{PHOEBE VII: Interferometric module}
\shortauthors{Brož et al.}

\title{Physics Of Eclipsing Binaries. VII. Interferometric module}

\correspondingauthor{Miroslav Brož}
\email{mira@sirrah.troja.mff.cuni.cz}

\author[0000-0003-2763-1411]{Miroslav Brož}
\affiliation{Charles University, Faculty of Mathematics and Physics, Institute of Astronomy, V Holešovičkách 2, CZ-18200 Praha 8}

\author[0000-0002-1913-0281]{Andrej Pr\v sa}
\affiliation{Villanova University, Dept.~of Astrophysics and Planetary Sciences, 800 E.\ Lancaster Ave, Villanova, PA 19085, USA}

\author[0000-0002-5442-8550]{Kyle E.~Conroy}
\affiliation{Villanova University, Dept.~of Astrophysics and Planetary Sciences, 800 E.\ Lancaster Ave, Villanova, PA 19085, USA}

\author[0000-0002-3985-4463]{Alžběta Oplištilová}
\affiliation{Charles University, Faculty of Mathematics and Physics, Institute of Astronomy, V Holešovičkách 2, CZ-18200 Praha 8}

\author[0000-0002-0504-6003]{Martin Horvat}
\affiliation{University of Ljubljana, Dept.~of Physics, Jadranska 19, SI-1000 Ljubljana, Slovenia}


\begin{abstract}
Interferometric measurements
are essential to constrain models of stellar systems,
by spatially resolving angular distances and diameters
well below the classical diffraction limit.
In this work,
we describe the interferometric module of Phoebe,
which could be used just for this purpose.
Since binaries in Phoebe are represented by a triangular mesh,
our complex model is based on the integration over triangles.
Consequently,
Roche distortion,
rotation,
non-synchronicity,
misalignment,
eclipses of components,
darkening,
reflection, or
irradiation
are all accurately accounted for.
For comparison purposes, we provide a simplified model,
where components are represented by circular disks.
The key point of our approach is a possibility of combination with other datasets
(light curves, radial velocities),
which allows to construct robust models of stellar systems.
\textcolor{red}{This draft refers to a development version of Phoebe, available at \myurl{https://github.com/miroslavbroz/phoebe2/tree/interferometry}.
It is not yet included in the official Phoebe repository!}
\end{abstract}


\section{Introduction}

Interferometric observations have been obtained by
optical, infrared, sub-mm, or radio instruments,
including
CHARA \citep{Brummelaar_2005ApJ...628..453T},
VLTI \citep{Abuter_2017A&A...602A..94G},
ALMA \citep{Brogan_2015ApJ...808L...3A},
EHT \citep{Akiyama_2019ApJ...875L...2E}, or
VLBI \citep{Ma_1998AJ....116..516M}.
They allow to detect details of the order of $\lambda/B$,
where
$\lambda$ is the wavelength,
$B$ the projected baseline,
instead of $1.22\,\lambda/D$,
where
$D$ is the aperture diameter
(i.e., the diffraction limit).

All of them observe fringes in the focal plane
(or equivalently, correlations in the correlator)
\begin{equation}
I(x', y') = I_0 \{1 + \mu\exp[2\pi{\rm i}(ux' + vy')]\}\,, \label{I_x__y_}
\end{equation}
where
$I$ is the measured intensity,
$I_0$ the mean intensity,
$x'$, $y'$ the angular focal-plane coordinates (in rad),
$(u, v) = \vec B/\lambda$ the projected baselines (in cycles per baseline),
or alternatively, `spatial frequencies'.

The appearance of fringes is determined by the complex visibility
\citep{Cittert_1934Phy.....1..201V,Zernike_1938Phy.....5..785Z}
\begin{equation}
\mu(u, v) = {1\over I_0} \int I(x, y)\exp[2\pi{\rm i}(ux + vy)] {\rm d}x{\rm d}y\,, \label{mu_u_v}
\end{equation}
which is the Fourier transform of the source intensity, in the sky plane.
Its real part $\Re\mu$ is the contrast of fringes,
while its complex part $\Im\mu$ is the shift of fringes,
corresponding to the phase shift in Eq.~(\ref{I_x__y_}).

In order to obtain $I(x, y)$,
the inverse Fourier transform should be applied as
\begin{equation}
I(x, y) = I_0 \int \mu(u, v)\exp[-2\pi{\rm i}(ux + vy)] {\rm d}u{\rm d}v\,,
\end{equation}
provided we have a complete information
(complex visibility, complete coverage).
However, this is rarely the case;
either, one has to interpolate and extrapolate $\mu(u, v)$,
or use a model of the source,
fitting an incomplete information
(real visibility, limited coverage).

Generally speaking, precise instruments require precise models.
Phoebe
\citep{Prsa_2016ApJS..227...29P,Horvat_2018ApJS..237...26H,Jones_2020ApJS..247...63J,Conroy_2020ApJS..250...34C}
is known to be very precise for stellar systems,
because it accounts for a number of effects,
Roche distortion,
rotation,
non-synchronicity,
misalignment,
eclipses of components,
including partial eclipses of triangles,
limb darkening,
gravity darkening,
reflection, or
irradiation.
Consequently,
it has a wide range of astrophysical applications, from
massive binaries,
close binaries,
contact systems,
dwarf novae, to
exoplanets.
Last, but not least,
it provides numerical methods for
estimation,
optimisation, and
sampling of parameters,
to perform their throughout statistical analysis.

Hereinafter, we describe how interferometry was incorporated to Phoebe,
namely,
methods in Sec.~\ref{methods} and
examples in Sec.~\ref{examples}.
In fact, we provide two models,
complex and simplified,
which allows for a comparison, a verification,
and most importantly,
an assessment, whether simplified models are applicable, or not.


\section{Methods}\label{methods}

\subsection{Complex model}\label{complex}

Formally, the complex model is a discrete alternative to Eq.~(\ref{mu_u_v})
\begin{equation}
\mu = {1\over L_{\rm tot}} \sum_i I_{{\rm pass},i} S_i \cos\theta_i f_i \exp\left[-2\pi{\rm i}(ux_i + vy_i)\right]\,,
\end{equation}
where
the sum is computed over triangles,
$x$, $y$ are the angular sky-plane positions (in rad), 
$u$, $v$ the projected baselines (in cycles per baseline),
$I_{\rm pass}$ the associated passband intensities,
which are limb darkened, gravity darkened,
i.e., the same as for light curve computations,
$S_i$ the surface areas,
$\theta_i$ the emergent angles,
$f_i$ the fractions of triangles,
which are visible.
For a normalisation of $\mu$ to 1,
the sum is divided by the total passband flux, $L_{\rm tot}$.

As an approximation,
we used a passband (V, R, I, J, H, K, \dots) intensity,
instead of monochromatic,
or narrow-band, corresponding to high spectral resolutions.
Nevertheless, the exact $\lambda$'s are used for $(u,v) = \vec B/\lambda$.
In other words,
an assumption that objects do not change much over the respective passband is used.
If needed, a dataset can be split
(e.g., H for PIONIER, K for GRAVITY),
or alternatively, narrower passbands can be defined by users.

In the future,
with a combination of interferometric and spectroscopic modules,
it will be possible to apply a more preceise weighting,
by the respective spectral-energy distributions.


\subsection{Simplified model}\label{simplified}

For circular disks,
both uniform or limb-darkened,
analytical solutions exist
\citep{Hanbury_1974MNRAS.167..475H}.
Since the Fourier transform is linear and a phase shift
is just a multiplication by a complex exponential,
it allows us to construct a simplified model,
suitable for separated, detached, non-eclipsing binaries.
It is inferior, but still useful for fast computations.

Let us start with the definitions,
of the Bessel function
\begin{equation}
J_{3/2}(x) = \sqrt{2\over\pi x} \left({\sin x\over x} - \cos x\right)\,,
\end{equation}
of the Planck function
\begin{equation}
B_\lambda = {2 h c^2\over \lambda^5}{1\over\exp\left({hc\over\lambda kT}\right)-1}\,,
\end{equation}
of the monochromatic luminosity
\begin{equation}
L_\lambda = \pi R^2 B_\lambda\,,
\end{equation}
of the argument
\begin{equation}
{\rm arg} = {\pi\phi B\over\lambda}\,,
\end{equation}
where
$\phi = 2R/d$ is the angular diameter, and
$d$ the system distance.
Then, the complex visibility
\begin{equation}
\mu = {1\over L_{\rm tot}} \sum_i L_{\lambda,i} \left({\alpha_i\over 2} + {\beta_i\over 3}\right)^{-1} \left[\alpha_i {J_1({\rm arg}_i)\over{\rm arg}_i} + \beta_i\sqrt{\pi\over 2} {J_{3/2}({\rm arg}_i)\over{\rm arg}_i^{3/2}}\right] \exp\left[-2\pi{\rm i}(ux_i + vy_i)\right]\,,
\end{equation}
where the sum is over components (not triangles),
$\alpha = 1-u_{\rm limb}$
$\beta = u_{\rm limb}$
are the coefficients related to the linear limb darkening law
\citep{Hanbury_1974MNRAS.167..475H}.


\subsection{Other interferometric observables}

Among observables, which are commonly recovered from fringes,
is the absolute value of visibility
\begin{equation}
V = |\mu|\,,
\end{equation}
or the squared visibility
\begin{equation}
V^2 = \mu\mu^*\,.
\end{equation}
Both of them relate to the contrast of fringes.

For a combination of three baselines, a triple product is defined as
\begin{equation}
T_3 = \mu(u_1, v_1) \mu(u_2, v_2) \mu(-(u_1+u_2), -(v_1+v_2))\,.
\end{equation}
Its argument $\arg T_3$ is called the closure phase.
Since the baselines above form a closed triangle,
it is a self-calibrating quantity,
which is not affected by seeing.
On the other hand,
the triple product amplitude $|T_3|$
behaves like a visibility
and needs a proper calibration.

In fact, a number of other observables is used in interferometry, e.g.,
estimators $C_1$, $C_2$ \citep{Roddier_1984JOpt...15..171R,Mourard_1994A&A...288..675M},
cross-spectrum $W_{12}$ \citep{Berio_1999JOSAA..16..872B,Berio_2001JOSAA..18..614B}, or
differential visibility $\Delta V$ \citep{Mourard_2009A&A...508.1073M}.
None of these is currently supported in Phoebe.


\subsection{Implementation notes}

We introduced
three new datasets in Phoebe,
VIS, CLO, T3.
The corresponding quantities
\verb|vises|, \verb|clos|, \verb|t3s|
were exposed to users (as `twigs').
For example,
\verb|vises = b.get_value('vises', dataset='vis01', context='model')|,
or alternatively,
\verb|vises = b['vises@vis01@phoebe01@latest@vis@model'].value|.
Further examples are available as Jupyter notebooks.
Standard OIFITS files
\citep{Pauls_2005PASP..117.1255P,Duvert_2017A&A...597A...8D}
could be input in a standard way,
with help of the oifits module.

All other features of Phoebe worked out of the box for these new datasets,
including computations of
distortions,
eclipses,
residuals,
$\chi^2$,
estimation,
optimisation,
sampling,
plotting,
etc.

Internally, we introduced \verb|original_index|, useful for
passing of $u$, $v$, $\lambda$ (in m) as input, and
passing of $V^2$, $\arg T_3$, $|T_3|$ as output,
which is three times faster compared to the previous approach.


\section{Examples} \label{examples}

Hereinafter,
we provide some examples
of the Phoebe interferometric module computations.
A comparison with other software packages,
e.g,
LITPRO \citep{TallonBosc_2008SPIE.7013E..1JT},
PMOIRED \citep{Merand_2022SPIE12183E..1NM}, or
Xitau \citep{Broz_2017ApJS..230...19B,Broz_2021A&A...653A..56B,Broz_2022A&A...657A..76B,Broz_2022A&A...666A..24B,Broz_2023A&A...676A..60B}
shows that our results are in agreement,
provided the respective approximations
(uniform disks, spherical components, \dots)
are fulfilled.

\subsection{Comparison of simplified vs. complex models}

\begin{figure}
\centering
\includegraphics[width=8cm]{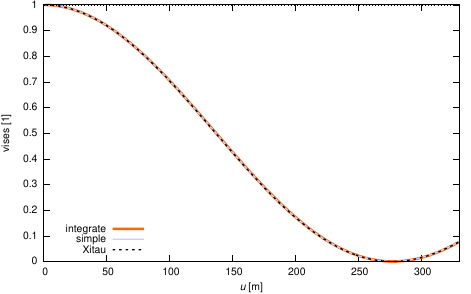}
\caption{
Basic example
of the squared visibility $|V|^2$ over baseline $u$ (in metres)
computed for
the default binary in Phoebe
(i.e.,
$P = 1\,{\rm d}$,
$a = 5.3\,R_\odot$,
$m_1 = m_2 \doteq 0.998813\,M_\odot$,
$R_1 = R_2 = 1\,R_\odot$).
The system was assumed to be at a distance of $100\,{\rm pc}$.
The wavelength $\lambda = 662.5\,{\rm nm}$,
the corresponding passband was Johnson~R.
Since the components are approximately spherical,
the complex model with integration over triangles (\color{orange}orange\color{black}),
the simple model with limb-darkened disks (\color{gray}gray\color{black}),
and the comparison with Xitau (black dotted)
all agree very well.
}
\label{test_phoebe3}
\end{figure}

The first example
is a computation of the squared visibility $|V|^2$
for the default binary in Phoebe.
It is detached, separated, with spherical components,
so the approximations of the simplified model are acceptable,
as verified by the complex model
(see Fig.~\ref{test_phoebe3}).
Of course, this is only true out of eclipses

\subsection{Closure phase}

\begin{figure}
\centering
\includegraphics[width=8cm]{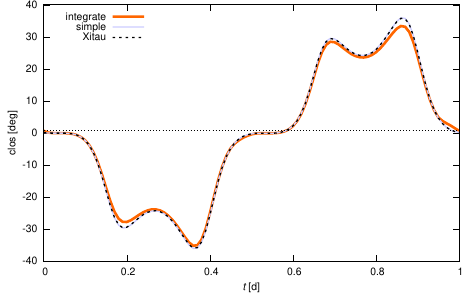} \\
\caption{
Closure phase example
computed for the default binary,
with different inclination
($i = 80^\circ$)
and different secondary temperature
($T_2 = 5000\,{\rm K}$).
In this situation,
the photocentre is moving,
and so is the closure phase
(${\rm arg}\,T_3$).
}
\label{test_phoebe5}
\end{figure}

The second example,
is a computation of the closure phase ${\rm arg}\,T_3$,
where we assumed a different inclination
($i = 80^\circ$)
and a different secondary temperature
($T_2 = 5000\,{\rm K}$),
in order to induce photocentre motions,
which is directly related to the closure phase
(Fig.~\ref{test_phoebe5}).
Here, we prefer to plot a temporal dependence
for fixed baselines ($u_1, v_1, u_2, v_2$),
instead of different baselines.
Nevertheless, the simple model is again in agreement with the complex model.

\subsection{Limb darkening}

\begin{figure}
\centering
\includegraphics[width=7cm]{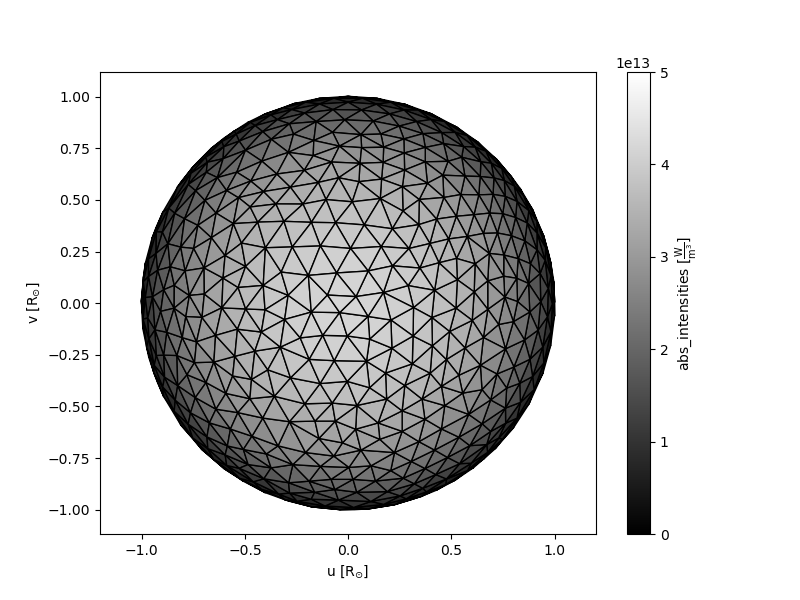}
\includegraphics[width=7.5cm]{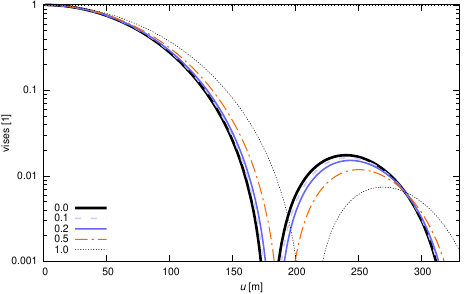}
\caption{
Limb-darkening example
computed for the default star in Phoebe
(i.e.,
$m = 1\,M_\odot$,
$R = 1\,R_\odot$);
observed at a distance of $10\,{\rm pc}$.
Limb darkening coefficients were set to manual, linear;
the values 
0.0,
0.1,
0.2,
0.5,
1.0
were tested.
The visibility changes as expected.
}
\label{test_interferometry4_limb}
\end{figure}

The third example
are computations for different limb darkening coefficients.
We assumed a linear law,
parametrized by a single number
(Fig.~\ref{test_interferometry4_limb}).
The visibility changes as expected;
if the limb is dark,
the apparent radius is small,
and the baseline needed to measure it is long.
Moreover, the second maximum of $|V|^2(B)$
is a known constraint for the limb darkening.
In Phoebe,
the limb darkening coefficients are tied to the atmospheres;
they are no longer considered as free parameters.
Instead, the effective temperature, gravity ($g = Gm/R^2$), or metallicity
would have to be adjusted to fit such interferometric observations.

\subsection{Rotation}

If geometry or distortion of stellar surfaces is more complex,
the simplified model is no longer applicable.
This is especially true for fast-rotating stars,
which first become ellipsoidal
and then subsequently approach the critical surface
with a 3:2 ratio.
Moreover, these stars exhibit uneven temperature distribution
\citep{vonZeipel_1924MNRAS..84..665V,Aufdenberg_2006ApJ...645..664A},
and --if inclined--
an asymmetric intensity distribution.

\begin{figure}
\centering
\includegraphics[width=7cm]{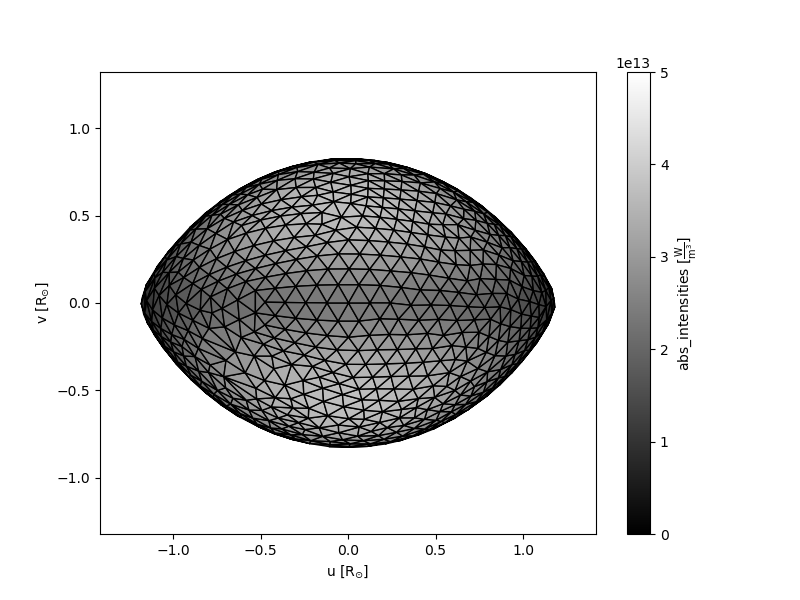}
\includegraphics[width=7.5cm]{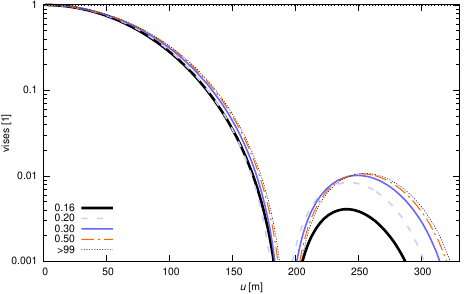}
\caption{
Rotation example
computed for the default star in Phoebe.
The rotation period values
0.16 (close to critical),
0.20,
0.30,
0.50, and
${>}99$~days
were tested.
The visibility changes as expected.
}
\label{test_interferometry6_rotation}
\end{figure}

We computed several models,
from the critical rotation rate to slow-rotating stars
(Fig.~\ref{test_interferometry6_rotation}).
The 3:2 ratio in two perpendicular directions
is probably the easiest measure of rotation,
however,
there are substantial differences among these models
at the second maximum of $|V|^2(B)$,
measurable at the per-cent level.

\subsection{Eclipses}

\begin{figure*}
\centering
\begin{tabular}{cc}
\includegraphics[width=6cm]{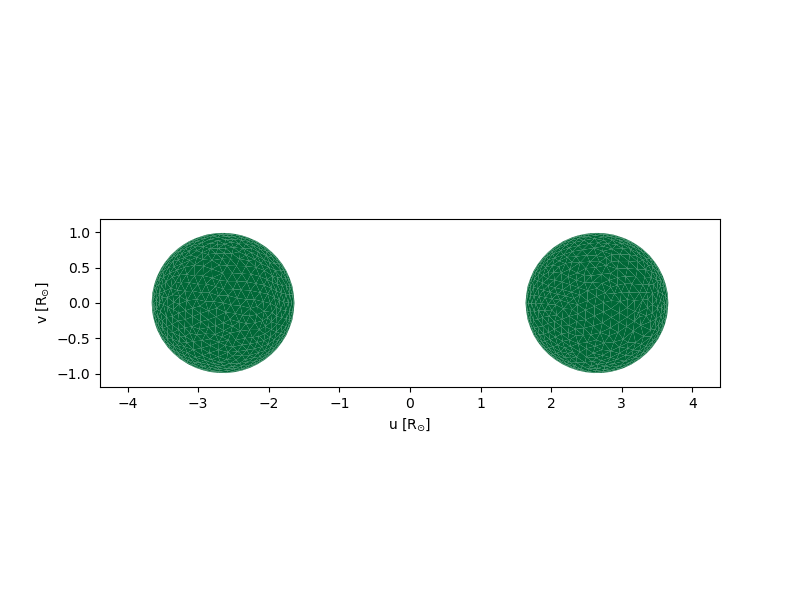} &
\includegraphics[width=5cm]{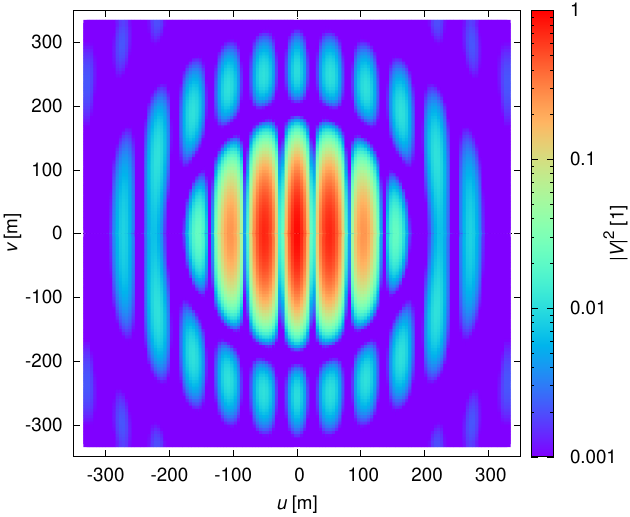} \\[-0.3cm]
\includegraphics[width=6cm]{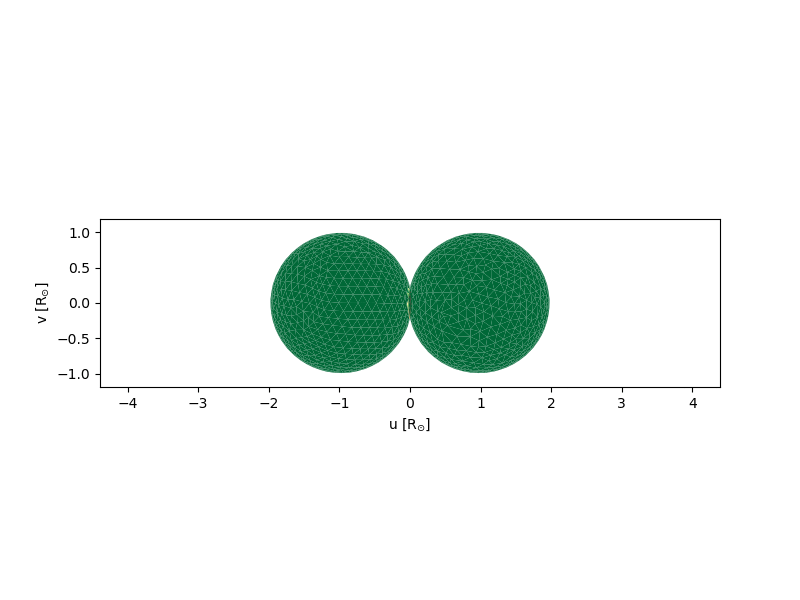} &
\includegraphics[width=5cm]{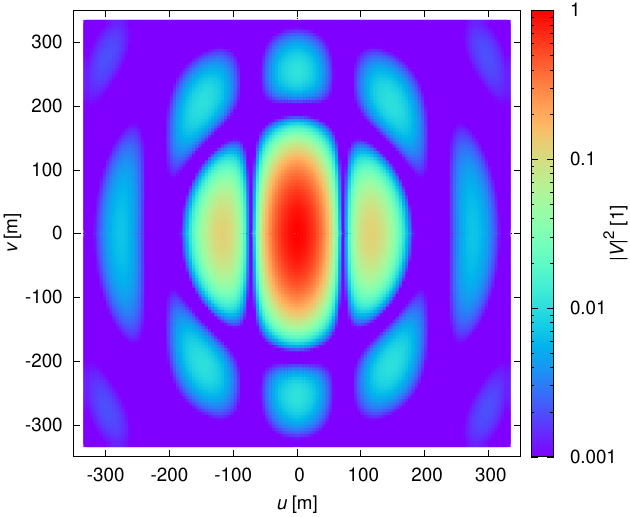} \\[-0.3cm]
\includegraphics[width=6cm]{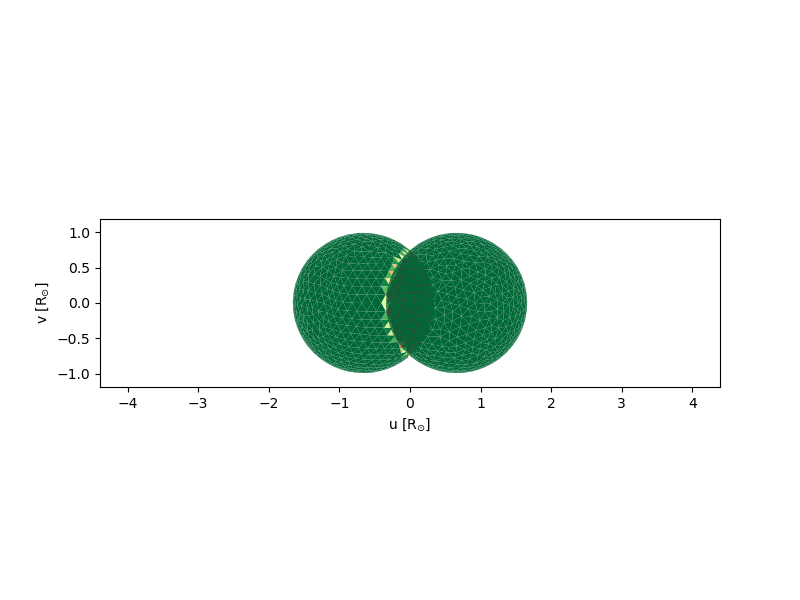} &
\includegraphics[width=5cm]{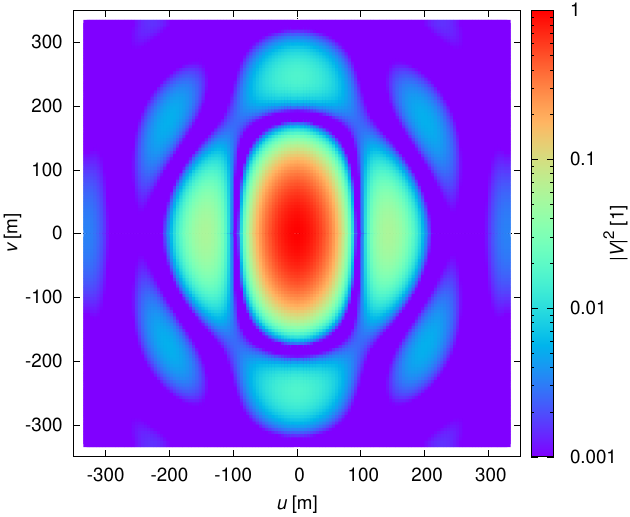} \\[-0.3cm]
\includegraphics[width=6cm]{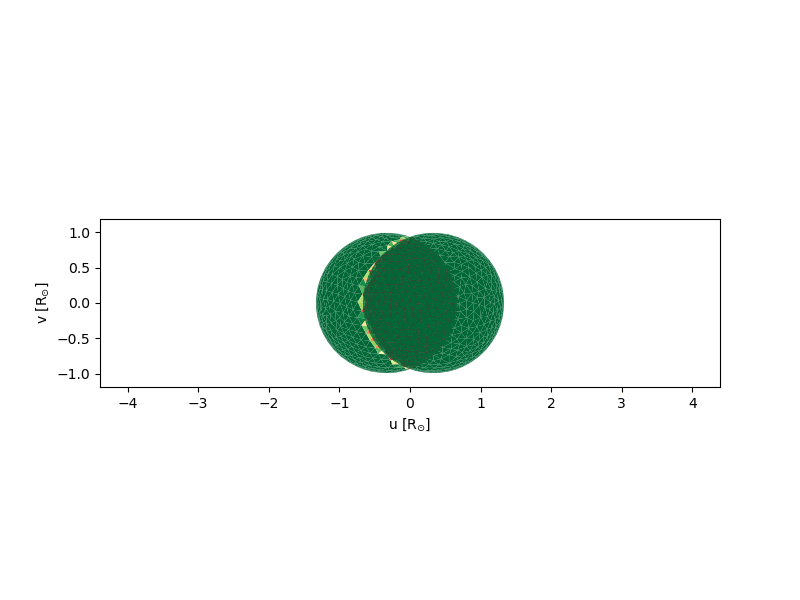} &
\includegraphics[width=5cm]{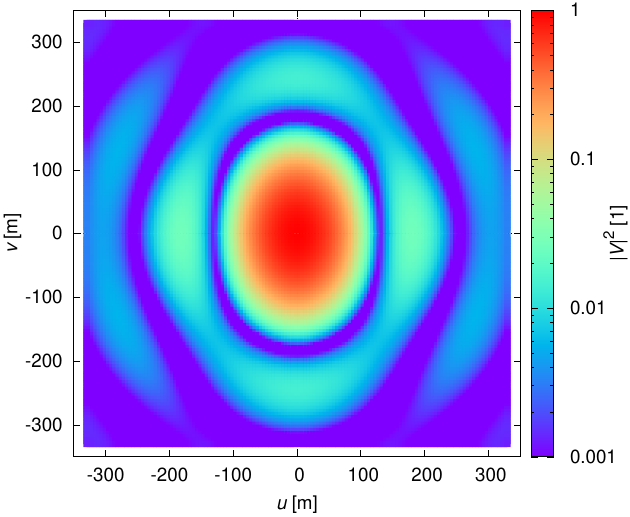} \\[-0.3cm]
\includegraphics[width=6cm]{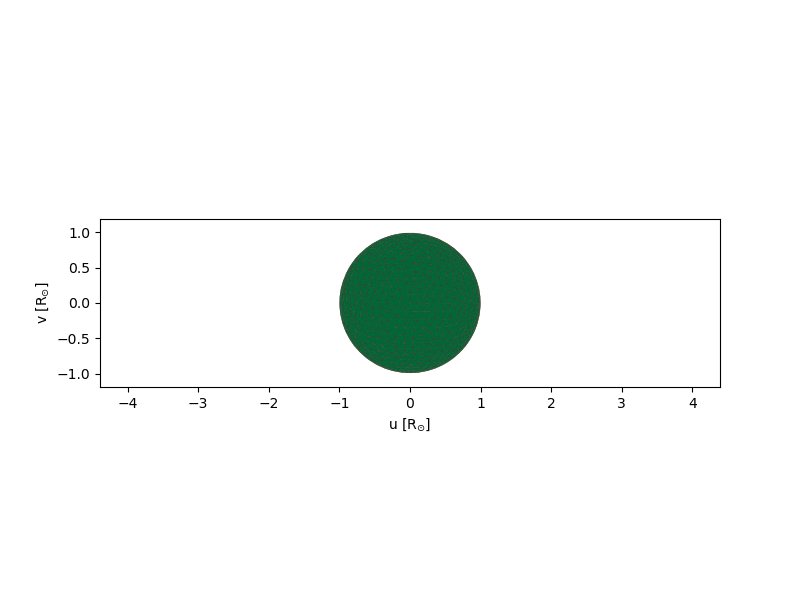} &
\includegraphics[width=5cm]{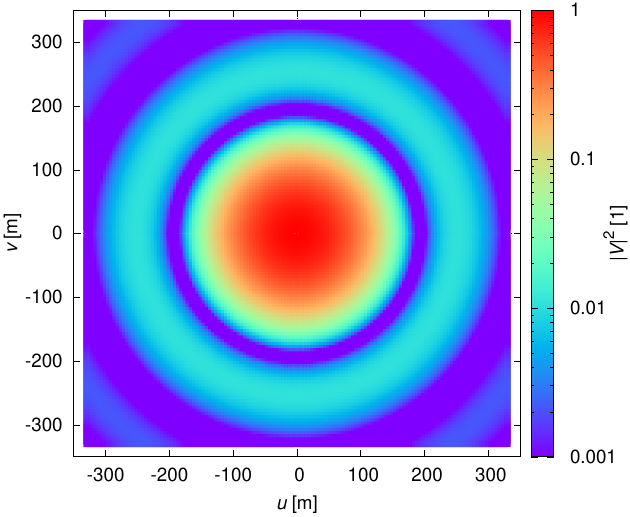} \\
\end{tabular}
\caption{
Eclipse example
computed for the default binary in Phoebe.
Left: Meshes with the fractions of triangles, which are visible.
Right: Squared visibility $|V|^2$ (colour),
as a function of the baselines $u$, $v$ (in metres).
In the course of the eclipse,
a pattern characteristic for a binary changes smoothly into a single-star.
}
\label{test_interferometry3_anim}
\end{figure*}

If eclipses occur in the course of interferometric observations,
the simplified model is again no longer applicable.
In Phoebe,
eclipses are computed precisely
\citep{Prsa_2016ApJS..227...29P}
and so are the interferometric observables.
We thus computed a series of models
for the default binary,
to demonstrate this capability
(Fig.~\ref{test_interferometry3_anim}).
In the course of the eclipse,
a pattern characteristic for a binary changes smoothly into a single-star,
and {\em vice versa\/}.
Nevertheless, it is true that interferometric observations
--if planned--
are planned out of eclipses,
in order to reach the maximal spatial resolution.

\subsection{Multiple systems}

\begin{figure}
\centering
\includegraphics[width=11cm]{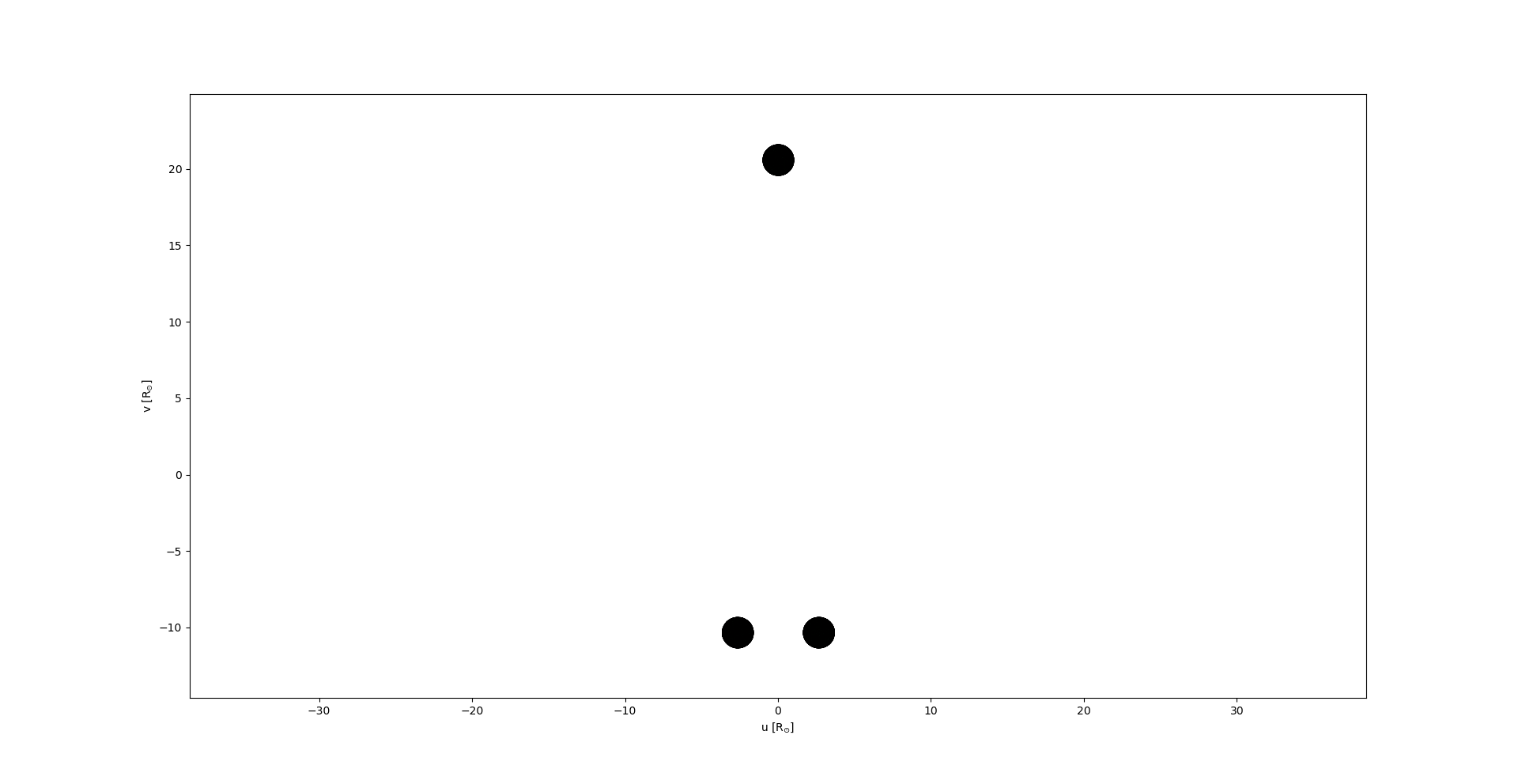} \\
\includegraphics[width=11cm]{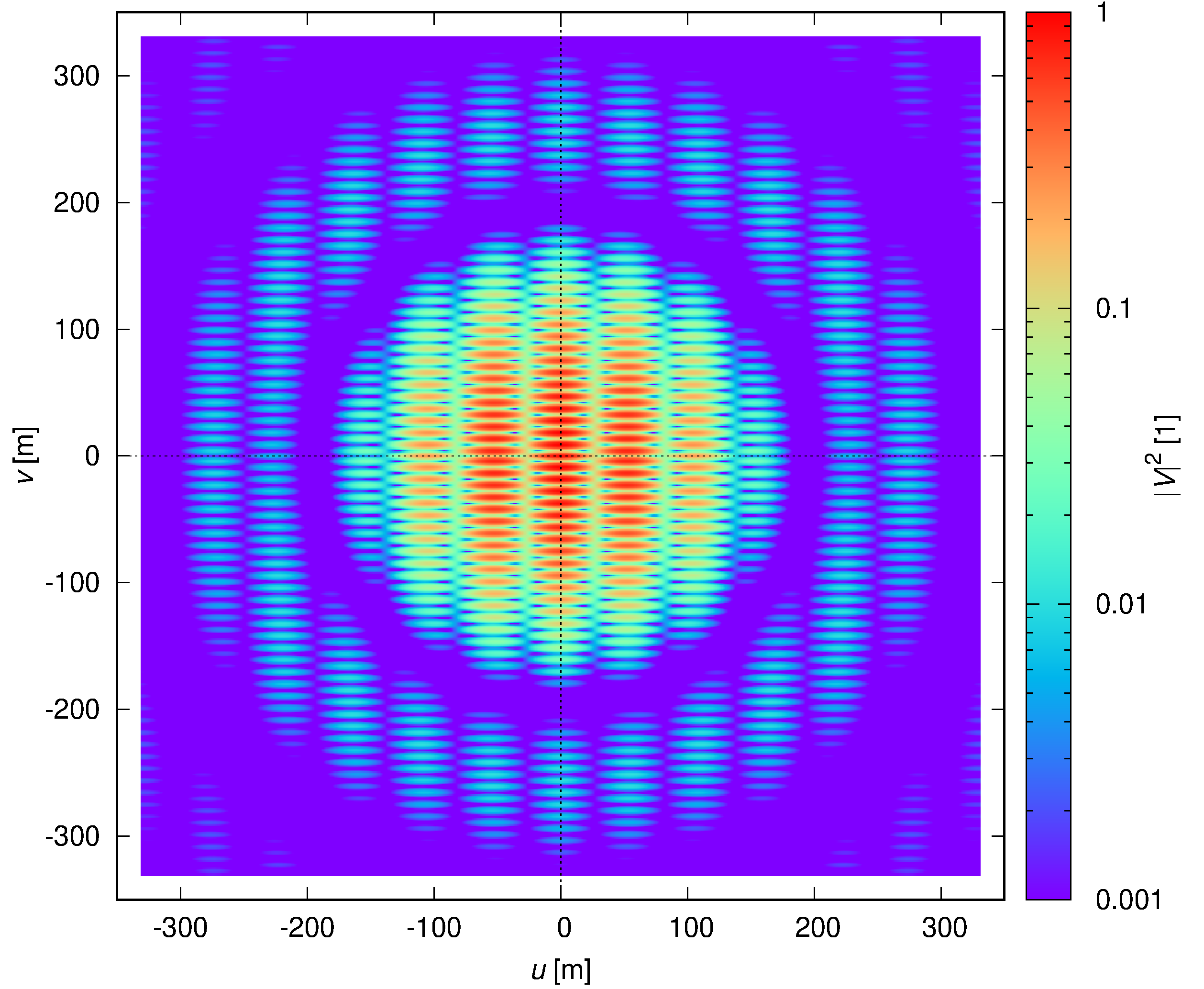}
\caption{
Triple system example
computed for the default triple in Phoebe
(i.e.,
$P_1 = 1\,{\rm d}$,
$P_2 = 10\,{\rm d}$,
$a_1 = 5.3\,R_\odot$,
$a_2 \doteq 30.994588\,R_\odot$,
$m_1 = m_2 \doteq 0.998813\,M_\odot$,
$m_3 = 1\,M_\odot$,
$R_1 = R_2 = 1\,R_\odot$).
A pattern is characteristic of two binaries
at different distances and at different orientations.
}
\label{test_interferometry5_triple}
\end{figure}

Finally,
multiple systems can be computed too,
with a development version of Phoebe.%
\footnote{\url{https://github.com/miroslavbroz/phoebe2/tree/interferometry}}
This is useful for O and B stars,
which are almost never single
and commonly more than double
\citep{Duchene_2013ARA&A..51..269D}.
We computed a model
for the default triple
to demonstrate this capability
(Fig.~\ref{test_interferometry5_triple}).
As expected,
the squared visibility $|V|^2(u, v)$ exhibits
a pattern characteristic of two binaries
at different distances and at different orientations.

Note this particular triple system a bit too compact
and the Keplerian dynamics is no longer sufficient
to describe its motion.
To avoid the respective systematics,
one should prefer the N-body dynamics, to account for
perturbations,
oblateness,
and relativistic effects
(using a development version of Phoebe).

\section{Conclusions}

We have described the interferometric module of Phoebe,
with a number of examples,
and how it could be used.
However, the key point is a combination of interferometric observations
with other 'orthogonal' observations
(light curves, radial velocities).
This should allow to construct robust models of stellar systems,
with a per-cent level of accuracy
\citep{Pietrzynski_2013Natur.495...76P}.

The first application will be to $\varepsilon$~Ori
and other Orion belt stars,
for which we have successfully obtained
interferometric measurements
by the VLTI/GRAVITY and PIONIER instruments
\citep{Oplistilova_2025}.





\begin{acknowledgements}
This work has been supported by the Czech Science Foundation through
grant 25-16507S (M. Brož).
\end{acknowledgements}

\bibliographystyle{apj}
\bibliography{paper1}

\end{document}